\documentstyle[aps,multicol]{revtex}

\begin{document}
%\draft
%\begin{center}
%{\Large \bf Observation of Stochastic Coherence in Coupled Map Lattices}
%\end{center}

%\vspace{2pt}

%\begin{center}
%Manojit Roy and R. E. Amritkar
%{\it Department of Physics,
%University of Pune, 
%Pune--411007, INDIA.}
%\end{center}

%\begin{title}
%Observation of Stochastic Coherence in Coupled Map Lattices
%\end{title}
%\author{ Manojit Roy and R. E. Amritkar }
%\begin{instit}
%Department of Physics, University of Pune, Pune--411007, INDIA.
%\end{instit}
%\receipt{ }

\title{ Observation of Stochastic Coherence in Coupled Map Lattices }
\author{ Manojit Roy and R. E. Amritkar \\
\it Department of Physics, University of Pune, Pune--411007, INDIA. }
\maketitle

%\vspace{5pt}

\begin{abstract}
Chaotic evolution of {\it structures} in Coupled map lattice driven by
identical noise on 
each site is studied  
(a structure is a group of neighbouring lattice--sites 
for whom values of dynamical variable follow certain predefined pattern).
Number of structures is seen to follow a {\it power--law decay} with
length of the structure for a given noise--strength.
An interesting phenomenon, which we call {\it stochastic coherence}, is reported
in which
a rise of {\it bell--shaped} type in abundance and lifetimes of
these structures is 
observed 
for a range of noise--strength values. 
\end{abstract}

%\vspace{15pt}

%\hspace{40pt}
%PACS number(s): 05.45.+b, 47.52.+j


\begin{multicols}{2}
%\newpage
\narrowtext

Coupled map lattice (CML) model has been observed to exhibit diverse
spatio--temporal patterns and structures \cite{OP}, and serves as a simple
model for experimental systems such as
Rayleigh--B\'enard convection, Taylor--Couette flow, B--Z reaction etc. \cite{RK}. 
A major 
interest is in understanding formation of structures, localised in both space 
and time, in turbulent fluid \cite{AKMH}.
Role of fluctuations in onset, selection and evolution of such patterns
and structures
has been studied in some detail \cite{MM,JK}. 
In this communication we report a novel phenomenon
observed in the dynamics of structures in a chaotically evolving one dimensional
CML driven by identical noise. By 
a {\it structure} we mean a group of neighbouring sites whose variable--values 
follow certain prespecified spatial pattern. 
Distribution of these structures during evolution of the 
lattice shows that their numbers exhibit {\it power--law decay} with length
of the
structure
for a given noise--strength, with an exponent which is a function of 
noise--strength.
It is observed that average length 
of these structures shows a bell--shaped curve with a characteristic peak,
as a 
function of noise--strength.
Similar behaviour is observed for average lifetime 
of these structures
during their evolution, within same range of noise values.
We call this phenomenon {\it stochastic coherence}.

We consider a one dimensional CML of the form
%\widetext
\begin{eqnarray}
x_{t+1}(i) & = & ( 1 - \varepsilon ) F(x_t(i)) \nonumber \\ && +
\;{\varepsilon \over 2 } 
\Big[ F(x_t(i-1)) + F(x_t(i+1))  \Big] 
+ \eta_t\;, \label{sc1} 
\end{eqnarray}
%\narrowtext
where $x_t(i)$, $i = 1, 2, \cdots, L$, is value of the variable located
at site $i$ at 
time $t$, $\eta_t$ is the 
additive noise, $\varepsilon$ is the (nearest neighbour) 
coupling strength, and $L$ is the size of the lattice. 
Logistic function $F(x) = \mu x (1 - x)$ is used as local dynamics 
governing nonlinear site--evolution with nonlinearity parameter $\mu$. We have 
used both open--boundary conditions
$x_t(0) = x_t(1),\;x_t(L+1) = x_t(L)$ , and periodic--boundary conditions 
$x_t(L+i) = x_t(i)$, for our system. 
For noise $\eta_t$ we have used 
a uniformly distributed random number bounded between $-W$ and $+W$,
where $W$ is defined as {\it noise--strength} parameter.

We define a {\it structure} as a region of space in which the dynamical
variables 
at sites within this region follow a predefined spatial pattern \cite{JA}.
To study 
coherence in
the system we choose a spatial pattern where the difference in the
values of the variables of neighbouring sites within the structure is
less than a
predefined small
positive number say $\delta$, i.e.,
$\vert x_t(i) - x_t(i\pm1) \vert \leq \delta$.
We call $\delta$ the
{\it structure parameter}. We look for such patterns, or coherent
structures, to appear in course of
evolution of the model given by Eq.~\ref{sc1}.

We now present our main observations.
Values of $\mu$, $\varepsilon$ and $L$ are chosen so that the resultant
dynamics of the
system is chaotic.
Coherent structures with length $< 3$ (sites) and 
lifetime $< 2$ (timesteps) are disregarded in our observation.
Values for $W$ are chosen within the range $[0,1]$.
Figure 1 shows plot (on log--log scale) of distribution of number $n(l)$
vs. length $l$ 
of structures for different values of $W$, with $\mu = 4$,
$\varepsilon = 0.6$, 
$\delta = 0.0001$ and $L = 10000$, and open--boundary conditions are used. 
Power--law nature of decay of $n(l)$ is 
clearly evident, which has a form
\begin{eqnarray}
n(l) \propto l^{-\alpha}\;, \label{sc2}
\end{eqnarray}
where $\alpha$ is power--law exponent.
This indicates that the system does not have any intrinsic
length scale.
It may be noted that in absence of noise ($W = 0$) the decay is  
manifestly exponential \cite{JA}.
Exponent $\alpha$ 
(i.e., slope of the log--log plot in Fig.~1) is seen to depend on
noise--strength $W$. 
This fact is corroborated in Fig.~2 
which shows plot of $\alpha$ vs. $W$ for the same parameter values as in 
Fig.~1. The exponent is exhibiting a clear minimum for values of $W$ 
around $0.6$. We define average length $\bar{l}$ of a structure as 
$\bar{l} = \sum{l\,n(l)} / \sum{n(l)}$. In Fig.~3 we plot variation of
$\bar{l}$ with 
$W$ for values of parameters as in Fig.~1.
The plot exhibits a {\it bell--shaped} nature within a fairly 
narrow range of $W$ around value $0.6$. It may be noted that the minimum
of $\alpha$
in Fig.~2 also occurs for $W$ quite close to this value. 

We call the phenomenon observed above {\it stochastic coherence}. This is
similar to
stochastic resonance phenomenon which shows a bell--shaped behaviour of 
temporal response as function of noise--strength \cite{BPSV}. 
However one 
may note that our system does not have any intrinsic length--scale, whereas
in 
stochastic resonance noise resonates with a given time--scale;
hence our use of word coherence rather than resonance \cite{JK1}. In
stochastic resonance
noise transfers energy to the system at a characteristic frequency,
whereas
in stochastic coherence noise {\it induces coherence} to the system. 
At this stage it is not clear to us as to how noise is inducing
coherence to the system. However, absence of any length--scale over 
entire range of noise seems to indicate existence of
weak self--organisation in the system induced by noise.

In this context it should be noted that if a uniform--deviate random number is
taken as site--variable, then the probability
of a site to belong to a structure of length $l$ is
$p_\delta(l) \approx l(2\delta)^{l-1}(1 - 2\delta)^2$,
where $\delta$ is the structure parameter 
introduced earlier. Hence number of such structures in a lattice of 
size $L$ is given by
$n_\delta(l) \approx L(2\delta)^{l-1}(1 - 2\delta)^2\;$.
This shows an exponential decay of number of 
such structures with length, which is to be contrasted with
the power--law form (\ref{sc2}) obtained for our lattice. 

It is possible to show how noise helps reducing instability of the structures,
thereby increasing their abundance. 
Let us consider
stability matrix $M$ of a homogeneous state \{$\cdots,x_t,$ $x_t,x_t,\cdots$\}
(this may be
thought of as a large structure with $\delta = 0$ for 
simplicity).
At time $t+1$ the matrix
takes the simple form $M_{t+1} = JF_t^\prime\;$, where $J$ stands for
the familiar tridiagonal matrix (with diagonal elements $1 - \varepsilon$ and
offdiagonal elements $\varepsilon/2$ on either sides) and
$F_t^\prime \equiv F^\prime (x_t) (\equiv dF/dx) = \mu (1 - 2 x_t)\;$.
After two timesteps,  we get the stability matrix for three time steps as 
$S_3 = M_{t+1}M_{t+2}M_{t+3} = J^3 F_t^\prime F_{t+1}^\prime F_{t+2}^\prime = 
J^3F_t^\prime \mu^2 [(1 - 2F_t)\{1 - 2 \mu F_t (1 - F_t)\} +
6 \mu \eta_t^2(1 - 2F_t) - 2 \eta_t\{1 + \mu (1 - 6F_t + 6F_t^2)\} - 4 \mu \eta_t^3
- 2 \eta_{t+1}(1 - 2F_t) + 4 \eta_t \eta_{t+1} ]\;$. For delta--correlated noise
with uniform distribution and zero mean (which is the case here), after averaging
the above
expression over noise--distribution one gets nonzero contribution due to noise only
from the term quadratic in $\eta\;$ :
\begin{eqnarray}
<S_3> = J^3 (F_t^\prime)^2 \mu \Big[1 - 2 \mu F_t (1 - F_t) +
6 \mu <\!\eta_t^2\!> \Big], \label{sc3}
\end{eqnarray}
where $<>$ denotes averaging over noise.
For our lattice with localised structures in backdrop of spatiotemporal
chaos we found the invariant density to be no longer symmetric, with larger
weightage
for values of variable above 0.5. Averaging
expression (\ref{sc3}) over this density makes the term $1 - 2 \mu F_t (1 - F_t)$
negative.
Adding positive contribution of noise to it results in reduction of eigenvalue
of the matrix $S_3$ and hence consequent reduction in instability of the state.

To study evolutionary aspects of these coherent
structures we obtained distribution of number $n(\tau)$ of structures vs. 
their lifetimes $\tau$ for different $W$.
This is shown in Fig.~4 (on log--linear scale). It exhibits decrease of 
$n(\tau)$ with $\tau$ with a 
{\it stretched exponential} type of decay having a form 
\begin{eqnarray}
n(\tau) \propto \exp\big(-({\rm const.})\tau^\beta\big)\;, \label{sc4}
\end{eqnarray}
where $\beta$ depends on $W$. 
We define average lifetime
$\bar{\tau}$ of a structure as
$\bar{\tau} = \sum{\tau\,n(\tau)} / \sum{n(\tau)}$. In Fig.~5 we plot
$\bar{\tau}$
vs. $W$ for parameter values as in Fig.~4. The graph also shows a
bell--shaped
feature with maximum for 
$W$ around $0.6$. This $W$ value is close to that corresponding to 
the extrema in figures 2 and 3.

In order to ascertain chaotic nature of system evolution we have calculated
lyapunov exponent spectra for our system. We find a number of lyapunov
exponents
to be positive, implying that the underlying evolution is chaotic. Maximum lyapunov 
exponent $\lambda_{\rm max}$ shows a minimum around noise--strength $0.6$ \cite{KI}.
We have studied variation of $\lambda$--spectrum with coupling strength
$\varepsilon$. 
$\lambda_{\rm max}$ remains fairly constant for $0.2 \leq \varepsilon \leq 0.8$
for entire range of $W$. Behaviour of $\bar{l}$ with $\varepsilon$ is also
investigated. It is found that
$\bar{l}$ increases monotonically with $\varepsilon$ for all $W$.
This fact is quite contradictory to what is expected
from $\lambda_{\rm max}$. This indicates that lyapunov exponent alone cannot be used
for proper characterisation of spatio--temporal features of the system.
We have also
obtained power spectrum of the time--series of the dynamical variable for a 
given site. The plot does not show any characteristic frequency and supports
chaotic behaviour.
In addition, we have calculated power spectrum of values $x_t(i)$,
$i = 1, \cdots, L$, at a
given time. Nature of the spectrum confirms our earlier observation that the
system does 
not have 
any intrinsic length--scale.
It may appear that
the behaviour of our system is similar to spatio--temporal
intermittency phenomenon \cite{CM}. However, our system does
not show any regular burst--type feature (indicative of intermittency) 
in time; as already noted, power spectrum of time--series
does not have any distinctive peak for entire range of noise--strength.
Thus what we observe is 
development of 
appreciable coherence induced by noise in the system undergoing  
{\it spatially intermittent and temporally chaotic} evolution.

On quite a few occasions the entire lattice is seen to evolve as a single
coherent
structure to within the structure parameter $\delta$. 
We have studied lifetime of these coherent structures. 
Full lattice coherence appears for $W$ above a value around 0.4.
Interesting thing is that occasionally this state 
is seen to persist for fairly long durations (at times as 
long as 200 timesteps or more), but eventually the coherence is seen to
break up.
This demonstrates that for our system synchronised state is not a 
stable attractor.

However, in rare instances our system has been found to get into 
an apparently synchronised state after a very large time (larger than $10^7$
steps).
This is obtained
because of finite accuracy of computation which cannot 
distinguish unstable synchronised state from stable one \cite{ASP}. As
demonstrated
in the following, for our system this type of behaviour is an artifact 
resulting due to
combination of finite size of the lattice and finite accuracy.
Let us denote by $\bar{T}$ the average of timesteps required for first 
occurrence of full lattice
coherence, the average taken over different initial conditions.
We have studied 
variation of
$\bar{T}$ 
with structure parameter $\delta$ for a fixed lattice--size $L$, which shows
a definitive power--law of the form $\bar{T_L} \propto \delta^{-\gamma}$,
where
$\gamma$
is power--law exponent.
This indicates that the synchronised state, i.e., full lattice coherent
structure with $\delta = 0$, cannot be reached.
We have also investigated $\bar{T}$ vs. $L$ variation for a fixed $\delta$. 
We see a power--law behaviour of form
\begin{eqnarray}
\bar{T_\delta} \propto L^\nu\;, \label{sc5}
\end{eqnarray}
where $\nu$ is power--law exponent.
Thus even full lattice coherence with finite $\delta$ cannot be achieved as 
$L \to \infty$.

We now show that existence of full--lattice coherence with nonzero $\delta$  
(distinct from synchronised state) 
in a finite lattice is essentially
a consequence of power--law variation (\ref{sc2}) of $n(l)$ with $l$.
From relation (\ref{sc2})
probability of a site to belong to a structure of length $l$ is 
$p_\delta(l) \propto l.l^{-\alpha} = l^{1-\alpha}$. Therefore, the
probability 
of a site to belong to a structure of length $\geq L$
is 
$P_0 \equiv P_\delta(\geq L) 
\propto \sum_L^\infty{l^{1-\alpha}} \approx \int_L^\infty{dl\,l^{1-\alpha}}
\propto
L^{2-\alpha}$ (for $\alpha > 2$, which 
is the case in the entire range of our observation as can be seen
from Fig.~2). 
The probability that for the first time a site belongs to 
a structure of length $\geq L$ at timestep $T$ 
is then 
$P_\delta(T) \propto (1 - P_0)^{T-1}P_0$. Therefore average of $T$ is
given as
$\bar{T}_\delta(\geq L) = \sum_{T=1}^\infty{T\,P_\delta(T)}
\propto (P_0)^{-1}$,
i.e.,
\begin{eqnarray}
\bar{T}_\delta(\geq L) \propto L^{\alpha-2}\;. \label{sc6}
\end{eqnarray}
This demonstrates that full--lattice coherence (with finite $\delta$) will be 
observed in a finite lattice. Fig.~2 tells us that $\alpha - 2 \approx 0.22$
for $W = 0.6$, whereas we obtain $\nu (\rm {relation(\ref{sc5})}) \approx 0.3$.
We believe the
discrepancy to be due to boundary corrections, since (\ref{sc6}) is obtained
for an infinite lattice whereas (\ref{sc5}) holds for finite lattices.

Above observations were also carried out for several values of
$\varepsilon$ ranging from $0.1$ to $0.9$, as well as
for nonlinearity parameter $\mu$ between $3.6$ and $4$. All the features 
remain essentially the same.
We have also observed similar behaviour with
periodic--boundary conditions for the lattice.

In conclusion, we have reported a new phenomenon observed in a chaotically  
evolving one--dimensional CML driven by identical noise, which we termed
{\it stochastic coherence}. It is observed that there is a phenomenal increase 
in abundance of coherent structures of all scales due to noise.
By considering stability matrix during three time steps, we have been able
to show that noise can reduce instability of these structures.
Distribution of these structures shows a power--law decay with length of the 
structure,
with an exponent which shows a minimum at some intermediate noise--strength. 
Average length
as well as average lifetime of these structures exhibit characteristic maxima
at a noise--strength quite close to previous value.
This bell--shaped feature is similar to that of 
stochastic resonance which is a temporal phenomenon.
However, we emphasise that
our system does not have any intrinsic length--scale, whereas stochastic 
resonance is associated with a particular time--scale. 
These observations demonstrate that noise can play a major role in formation
as well
as in evolutionary dynamics of structures in spatially extended systems.

\vspace{.3 in}
\noindent
The authors thank H. Cerdeira and P. M. Gade for useful discussions.
One of the authors (M.R.) acknowledges University Grants Commission (India)
and the
other (R.E.A.) acknowledges Department of Science and Technology (India) for 
financial assistance.

\newpage

\begin{center}

{\bf FIGURE CAPTIONS}

\end{center}
\vspace{0.2 in}

\begin{itemize}
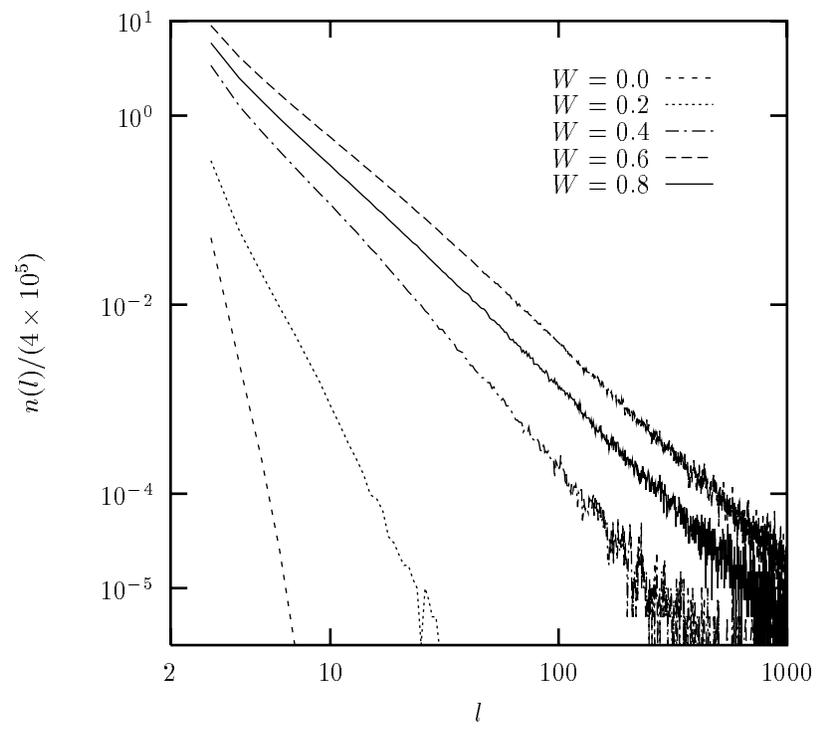
\item[Fig.~1.] Plot of variation of number $n(l)$ of structures 
with length $l$ for 
a lattice with size $L = 10000$, for different values of noise--strength $W$ as 
indicated.
Parameters chosen are coupling strength parameter
$\varepsilon = 0.6$, structure parameter $\delta = 0.0001$, and nonlinearity
parameter
$\mu = 4$. Open--boundary conditions are used. Data are obtained for 100000 
iterates per initial condition and 4 initial conditions.

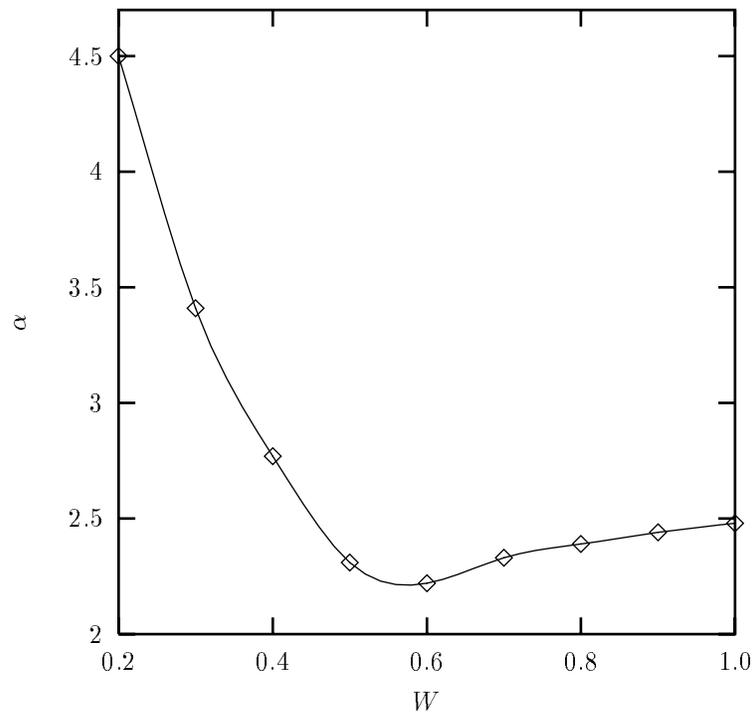
\item[Fig.~2.] Variation of exponent $\alpha$ (relation (\ref{sc2})) with
noise--strength 
$W$ plotted 
for parameter values as in Fig.~1.

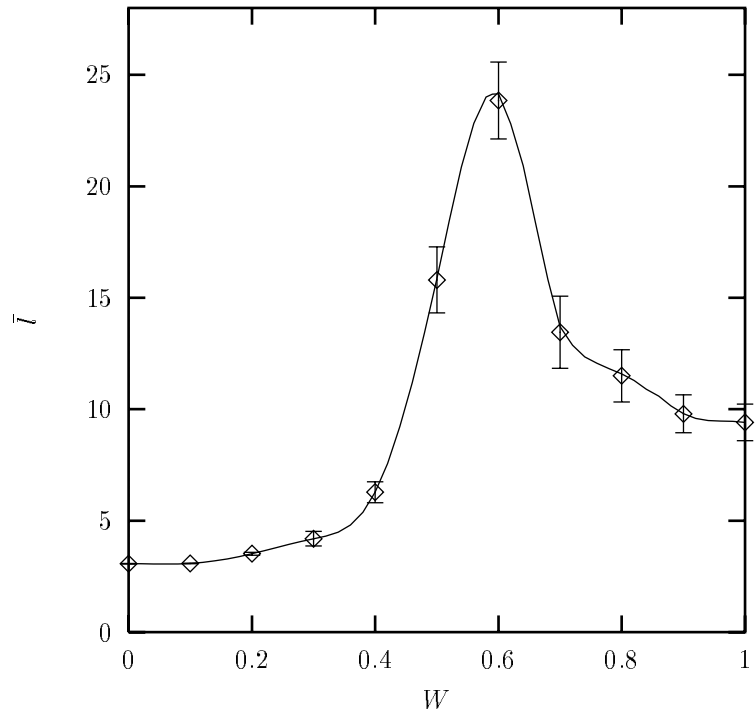
\item[Fig.~3.] Plot of variation of average length $\bar{l}$ of structure with 
noise--strength $W$, with parameters as stated in Fig.~1.

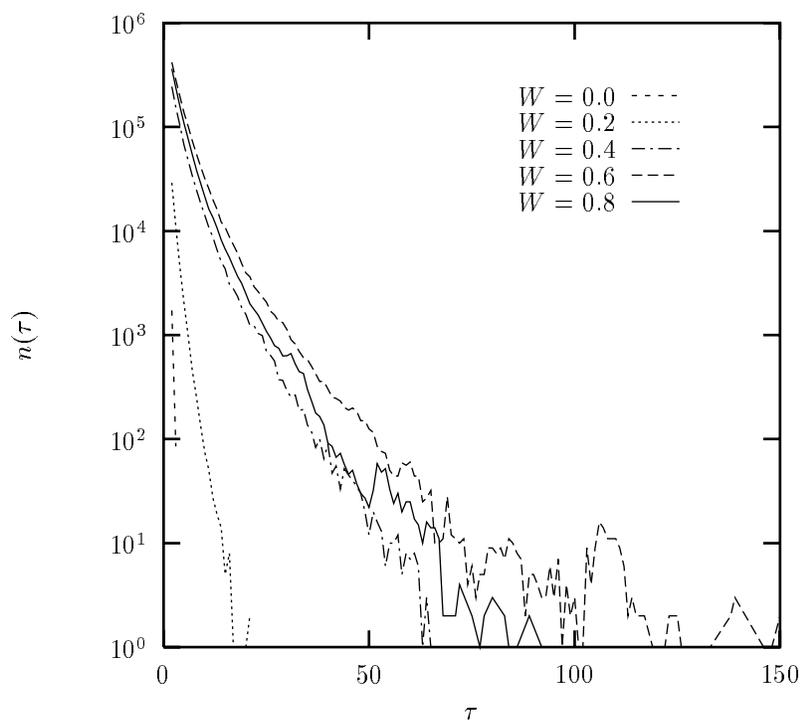
\item[Fig.~4.] Plot of variation of number $n(\tau)$ of structures
with lifetime $\tau$, with conditions as in Fig.~1.

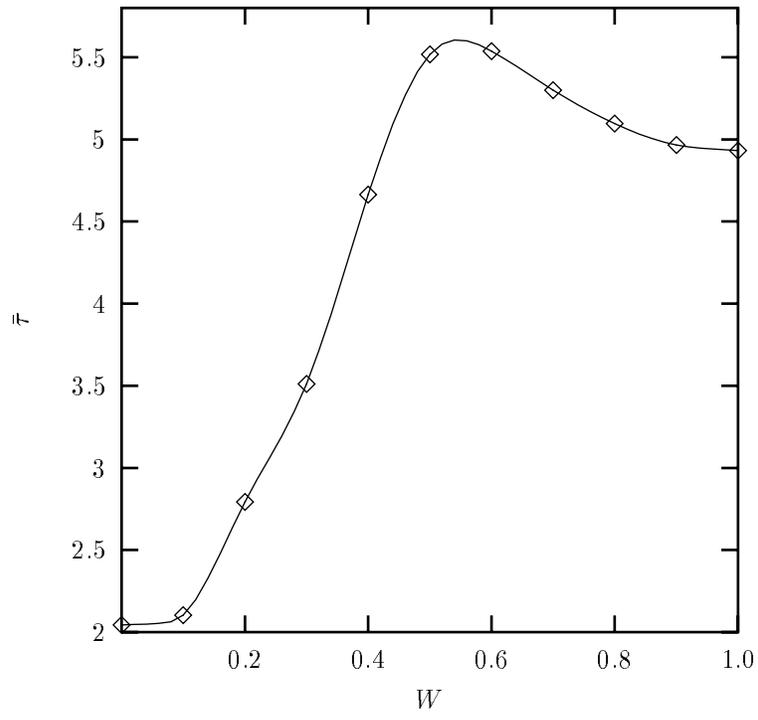
\item[Fig.~5.] Variation of average lifetime $\bar{\tau}$ of structures with 
noise--strength
$W$ shown for parameters as stated earlier.
\end{itemize}

\end{multicols}

\begin{thebibliography}{99}
%\begin{references}{99}
\bibitem{OP} Y. Oono and S. Puri, Phys. Rev. Lett. {\bf 58}, 836 (1986);
G. L. Oppo and R. Kapral, Phys. Rev. A {\bf 36}, 5820 (1987);
P. M. Gade and R. E. Amritkar, Phys. Rev. E {\bf 47}, 143 (1993);
R. E. Amritkar and P. M. Gade, Phys. Rev. Lett. {\bf 70},
3408 (1993);
{\it Theory and applications of coupled map lattices}
(edited by
K. Kaneko), John Wiley and Sons, Chichester, England (1993);
Qu Zhilin and Hu Gang, Phys. Rev. E {\bf 49}, 1099 (1994);
S. Raghavachari and James A. Glazier, Phys. Rev. Lett. {\bf 74},
3297 (1995).
\bibitem{RK} R. Kapral, in {\it Theory and applications of coupled map
lattices} (edited by K. Kaneko), p. 135;
M. C. Cross and P. C. Hohenberg, Rev. Mod. Phys. {\bf 65},
851 (1993).
\bibitem{AKMH} A. K. M. F. Hussain, Phys. Fluids {\bf 26}, 2816 (1983).
{\it Topological Fluid Mechanics} (edited by H. K. Moffatt
et. al.),
proceedings of the IUTAM symposium, 13--18 August 1989, Cambridge University 
Press (1990);
{\it Turbulence And Stochastic Processes: 
Kolmogorov's Ideas 50
Years On} (edited by J. C. R. Hunt et. al.), The Royal Society, London (1991).
\bibitem{MM} {\it Noise in Dynamical Systems} (edited by F. Moss and
P. V. E. McClintock), Cambridge University Press, Cambridge, U.K. (1989);
P. C. Hohenberg and J. B. Swift, Phys. Rev. A {\bf 46}, 4773
(1992); X--G Wu and R. Kapral, J. Chem. Phys. {\bf 100}, 1 (1994); 
S. Fahy and D. R. Hamann, Phys. Rev. Lett. {\bf 69}, 761 (1992);
D. A. Kurtze, Phys. Rev. Lett. {\bf 77}, 63 (1996); A. Maritan and J. R.
Banavar, Phys. Rev. Lett. {\bf 72}, 1451 (1994). Effect of noise on evolution
of patterns is also investigated by Kapral et al. in \cite{OP}.
\bibitem{JK} P. Jung and G. Mayer--Kress, Phys. Rev. Lett. {\bf 74}, 2130
(1995).
\bibitem{JA} J. K. John and R. E. Amritkar, Phys. Rev. E {\bf 51},
5103 (1995).
\bibitem{BPSV} R. Benzi, G. Parisi, A. Sutera and A. Vulpiani, Tellus 34,
10 (1982);
B. McNamara, K. Wiesenfeld and R. Roy, Phys. Rev. Lett. {\bf 60},
2626 (1988);
R. F. Fox and Yan--nan Lu, Phys. Rev. E {\bf 48}, 3390 (1993).
\bibitem{JK1} Existence of spatio--temporal stochastic resonance in 
extended system is observed leading to enhancement of {\it temporal
characteristics} of the system due to a Gaussian noise, as reported in
\cite{JK}.
\bibitem{KI} K. Matsumoto and I. Tsuda,
J. Stat. Phys. {\bf 31}, 87 (1983); {\bf 34}, 111 (1984).
\bibitem{CM} H. Chat\'e and P. Manneville, Phys. Rev. Lett. {\bf 58},
112 (1987); Physica D {\bf 32}, 409 (1988).
\bibitem{ASP} A. S. Pikovsky, Phys. Rev. Lett. {\bf 73}, 2931 (1994).
%\end{references}
\end{thebibliography}
\end{document}